# Orbital Evolution of the P/1996 R2 and P/1996 N2 Objects

## S. I. Ipatov* and G. J. Hahn**

*\*Keldysh Institute of Applied Mathematics, Russian Academy of Sciences, Miusskaya pl. 4, Moscow 125047, Russia*
*\*\*Institute of Planetary Sciences, Rudower Chausse 5, Berlin 12489, Germany*
Received October 20, 1998

**Abstract**—A numerical integration of the equations of motion of the Sun–planets–an object system is used to study the evolution of orbits close to the orbit of the P/1996 R2 object, which is a Jupiter-crossing object, and to the asteroidal orbit of the P/1996 N2 object, which, at the moment of its detection, had a tail similar to a cometary one. Small variations in the initial data considerably affect the evolution of orbits close to that of the P/1996 R2 object. The time elapsed up to the ejection of the object into a hyperbolic orbit varied from $3 \times 10^4$ to $2.7 \times 10^7$ yr. Some objects were in resonances with Jupiter and Saturn for a long time. For about 20% of the runs, objects reached the Earth's orbit during evolution. Orbital elements of the P/1996 N2 object changed quasi-periodically over the considered time span of 200 Myr. Variations in the semimajor axis, eccentricity, and inclination were equal to 0.04 AU, 0.11°, and 3.5°, respectively.

## VARIANTS OF RUNS

The P/1996 R2 (Lagerkvist) object, discovered by Lagerkvist in 1996, is one of the Jupiter-crossing objects. The P/1996 N2 (Elst–Pizarro) object has a typical nonresonant stable asteroidal orbit, but at the moment of its detection it had a tail which looked like that of a comet. It is considered that this tail appeared due to the collision of the object with a smaller asteroid. This object was identified as asteroid 1979 OW7. In some conference abstracts, Ipatov and Hahn (1997a–c) considered the evolution of orbits close to the orbits of these two objects. In the present paper, the results of these investigations are presented in greater detail.

In each series of calculations of the evolution of orbits close to the orbits of P/1996 R2 (Lagerkvist) and P/1996 N2 (Elst–Pizarro), we considered the "basic" orbit and 10 or 11 "neighboring" orbits. The difference between the elements of the basic orbit and a neighboring orbit was about the difference between the orbital elements obtained on the basis of different numbers of measurements.

For the P/1996 R2 object in the first series of calculations, along with the basic orbit ($a = 3.7905223$ AU, $e = 0.3126849$, $i = 2.60367°$, $\Omega = 40.27673°$, and $\omega = 334.41016°$), we assumed that the object passes its perihelion on January 20, 1997, 20.38224 TT. We examined 10 orbits, for which values of the semimajor axis $a$, the eccentricity $e$, the inclination $i$, the ascending-node longitude $\Omega$, and the perihelion argument $\omega$ varied by ±0.02 AU, ±0.008°, ±0.006°, ±0.1°, and ±1°, respectively (only one orbital element was varied for each orbit). For the second series of runs, these values were ±0.004 AU, ±0.0016°, ±0.0012°, ±0.02°, and ±0.23°, respectively, and were close to the difference in the orbital elements of two basic orbits obtained on the basis of 80 (September 11–October 9, 1996) and 90 (September 11–October 16, 1996) observations. For the second basic orbit, we used $a = 3.7863247$ AU, $e = 0.3111012$, $i = 2.60485°$, $\Omega = 40.25683°$, and $\omega = 334.18825°$ (the object passes its perihelion on January 19, 1997, 19.70954 TT). For the twelfth test object in each series, we considered the basic orbit from another series, but the date of the perihelion passage was identical for all runs in the entire series.

For P/1996 N2, the values of $a$, $e$, $i$, $\Omega$, and $\omega$ for "neighboring" orbits differ from analogous values for the "basic" orbit by ±0.005 AU, ±0.015°, ±0.05°, ±2°, and ±2°, respectively. These differences were greater by about a factor of 5 than the differences in the elements obtained on the basis of 15 (July 14–August 21, 1996) and 29 (from July 21, 1979 until August 21, 1996) observations. For the basic orbit we used: $a = 3.15619$ AU, $e = 0.16749$, $i = 1.38424°$, $\Omega = 160.266°$, and $\omega = 133.295°$ at JDT 2 450 200.5 (April 27, 1996). At that time the mean anomaly was $M \approx 1.52°$.

The integration was carried out with the use of the BULSTO integrator developed by Bulirsh and Stoer (1966) over the time span $T \pm 2$ Myr (into the past and into the future) for the P/1996 R2 object and over the time span $T = \pm 1$ Myr for the P/1996 N2 object. Using the faster (but less accurate) RMVS3 integrator from the SWIFT integration package worked out by Levison and Duncan (1994), we considered larger time intervals. The initial integration step for the RMVS3 integrator was taken to be 30 days. The relative accuracy of the integration step in Bulirsh and Stoer's method was

Results of the simulation of the evolution of orbits close to the orbit of the P/1996 R2 object

| a. Lifetimes $T_h$ (Myr). The BULSTO method of integration | | | | | | | | | | | |
|---|---|---|---|---|---|---|---|---|---|---|---|
| <−2. | −1.03 | −0.94 | −0.53 | −0.42 | −0.40 | −0.36 | −0.30 | −0.26 | −0.12 | −0.092 | −0.029 |
| 0.046 | 0.27 | 0.29 | 0.40 | 0.52 | 0.55 | 0.59 | 0.60 | 1.24 | 1.46 | 1.65 | 1.92 |
| <−2. | −1.23 | −1.15 | −0.89 | −0.85 | −0.80 | −0.79 | −0.77 | −0.48 | −0.42 | −0.31 | −0.14 |
| 0.051 | 0.061 | 0.086 | 0.172 | 0.174 | 0.21 | 0.22 | 0.23 | 0.26 | 0.90 | 1.73 | >2. |

| b. Lifetimes $T_h$ (Myr). The RMVS3 method of integration | | | | | | | | | | | |
|---|---|---|---|---|---|---|---|---|---|---|---|
| −9.26 | −7.57 | −2.5 | −1.08 | −0.54 | −0.47 | −0.43 | −0.41 | −0.31 | −0.14 | −0.083 | −0.070 |
| 0.067 | 0.10 | 0.12 | 0.17 | 0.22 | 0.23 | 0.27 | 0.37 | 0.56 | 1.20 | 1.24 | 4.15 |
| −4.64 | −0.65 | −0.64 | −0.42 | −0.39 | −0.26 | −0.22 | −0.19 | −0.15 | −0.14 | −0.050 | −0.015 |
| 0.048 | 0.052 | 0.075 | 0.093 | 0.12 | 0.26 | 0.74 | 0.80 | 0.95 | 1.18 | 1.47 | 27.06 |

| c. Minimum distances $q_{min}$ from the Sun (AU). The BULSTO method of integration. Integration to the past | | | | | | | | | | | |
|---|---|---|---|---|---|---|---|---|---|---|---|
| 1.09 | 1.25 | 1.39 | 1.42 | 1.54 | 1.73 | 1.75 | 2.47 | 2.60 | 2.60 | 2.60 | 2.60 |
| 0.002 | 0.30 | 1.05 | 1.12 | 1.23 | 1.32 | 1.49 | 2.11 | 2.29 | 2.60 | 2.60 | 2.60 |

| d. Minimum distances $q_{min}$ from the Sun (AU). The BULSTO method of integration. Integration to the future | | | | | | | | | | | |
|---|---|---|---|---|---|---|---|---|---|---|---|
| 0.63 | 0.72 | 0.81 | 0.91 | 0.92 | 1.05 | 1.57 | 1.90 | 1.91 | 2.06 | 2.15 | 2.15 |
| 0.50 | 0.68 | 0.98 | 1.49 | 1.78 | 1.87 | 1.95 | 1.98 | 2.00 | 2.02 | 2.15 | 2.17 |

| e. Minimum distances $q_{min}$ from the Sun (AU). The RMVS3 method of integration. Integration to the past | | | | | | | | | | | |
|---|---|---|---|---|---|---|---|---|---|---|---|
| 0.26 | 1.71 | 2.15 | 2.19 | 2.25 | 2.37 | 2.47 | 2.50 | 2.57 | 2.60 | 2.60 | 2.60 |
| 1.65 | 1.87 | 1.89 | 2.13 | 2.26 | 2.60 | 2.60 | 2.60 | 2.60 | 2.60 | 2.60 | 2.60 |

| f. Minimum distances $q_{min}$ from the Sun (AU). The RMVS3 method of integration. Integration to the future | | | | | | | | | | | |
|---|---|---|---|---|---|---|---|---|---|---|---|
| 0.98 | 1.17 | 1.21 | 1.31 | 1.46 | 1.93 | 2.21 | 2.53 | 2.57 | 2.60 | 2.60 | 2.60 |
| 0.71 | 0.83 | 1.41 | 1.42 | 1.94 | 1.94 | 1.99 | 2.27 | 2.60 | 2.60 | 2.60 | 2.60 |

set to be $10^{-9}$–$10^{-8}$. The influence of all planets, $n_{pl} = 9$, was taken into account in most runs. In the study of the orbital evolution of P/1996 N2 with the use of the RMVS3 integrator, the influence of Mercury was ignored, however, ($n_{pl} = 8$), because taking it into account introduced larger errors than neglecting this factor. Vashkov'yak, who reviewed this paper, noted that this fact is probably connected with a noticeably smaller step required for the numerical integration than that for the system without Mercury. For the P/1996 R2 object, time steps for constructing the plots were taken to be 100 yr for the BULSTO integrator and 500–1000 yr for the RMVS3 integrator. Using the RMVS3 integrator, we traced the evolution of the orbit to the time of ejection of the object into a hyperbolic orbit (up to 27 Myr). For P/1996 N2 and the BULSTO integrator, the orbital elements were computed with the step $\Delta t = 200$ yr for $T = \pm 1$ Myr and with the step $\Delta t = 10$ yr for $T = \pm 0.1$ Myr. For $T = \pm 20$ Myr, we used the RMVS3 integrator and took $\Delta t = 1000$ yr. For the basic orbit of this object, integration was also carried out for the interval $T = 200$ Myr into the future.

Using the BULSTO integrator, we showed that for $T = 0.1$ Myr, the differences $\Delta e = e_{max} - e_{min}$ and $\Delta i = i_{max} - i_{min}$ obtained for $n_{pl} = 9$ and $n_{pl} = 8$ are distinguished by only 0.000015° and 0.00004°, respectively, where "max" and "min" indicate the maximum and minimum values of the orbital elements over the time span under consideration. For an integration step of 30 days, the accuracy of integrations achieved with the RMVS3 integrator (in comparison with the results obtained with the BULSTO integrator) was better for $n_{pl} = 8$ than for $n_{pl} = 9$. For example, for $T = 0.1$ Myr, the difference in the $\Delta e$ values obtained with the RMVS3 and BULSTO integrators was ~0.002 for $n_{pl} = 9$ and ~0.0005 for $n_{pl} = 8$. For the differences in the $\Delta i$ values, with the same $n_{pl}$ values, we had 0.0014° and 0.0003°, respectively.

(a)

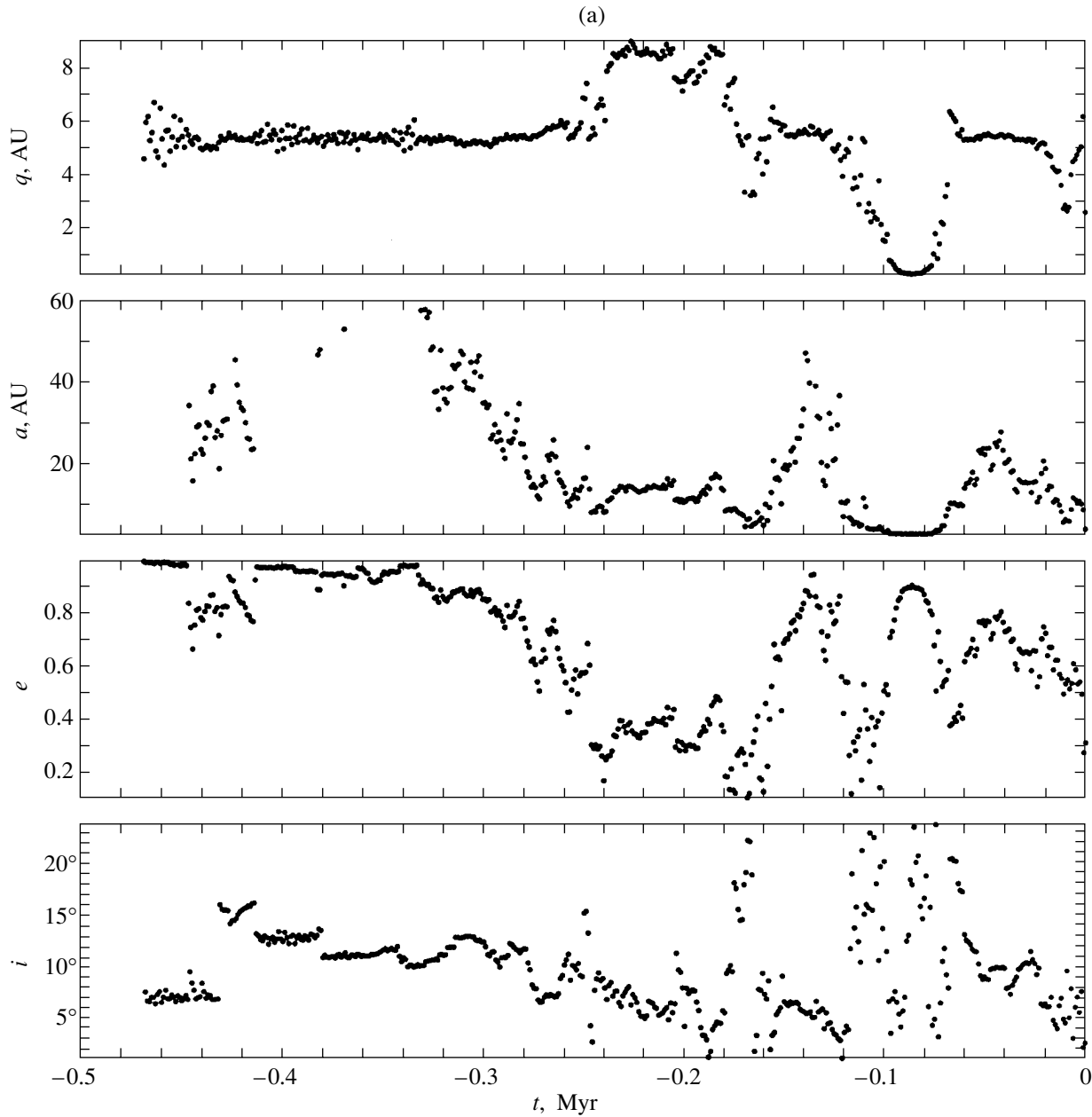

**Fig. 1.** Time variations (Myr) in the semimajor axis $a$ (AU), eccentricity $e$, perihelion distance $q = a(1 - e)$ (AU), inclination $i$ (in degrees) of a body, the difference $\Delta\Omega = \Omega - \Omega_J$ in the ascending-node longitudes of the body and Jupiter, the perihelion argument $\omega$, and the difference $\Delta\pi = \pi - \pi_J$ in the longitudes of the perihelion of the body and Jupiter (all angles are shown in degrees). For all runs, the initial orbits are close to the orbit of the P/1996 R2 object. The results are obtained by numerical integration of the system (the Sun, planets, an object) with the use of the RMVS3 integrator (a–c) and the method by Bulirsh and Stoer (d–j).

## EVOLUTION OF ORBITS CLOSE TO THE ORBIT OF THE P/1996 R2 OBJECT

Let us consider the results of the study of the evolution of orbits close to the orbit of the P/1996 R2 object. It was established that in 1990 (at JDT 2 448 061.45) a close encounter (0.21 AU) of P/1996 R2 with Jupiter occurred. The duration of strong variations in the elements of the heliocentric orbit was 4 yr. The next close encounter with Jupiter (0.57 AU) will take place in 2052 (at JDT 2 470 864.6). Only in 1996 did the minimum distance of P/1996 R2 from the Earth become less than 2 AU.

In the runs considered above, the time $T_h$ elapsed until the ejection of the object into a hyperbolic orbit varied from 0.05 to 27 Myr for integration to the future, and from –0.03 to –9.26 Myr for integration to the past (table a, b). With the use of the BULSTO integrator, the median lifetime for two series of runs was found to be

(b)

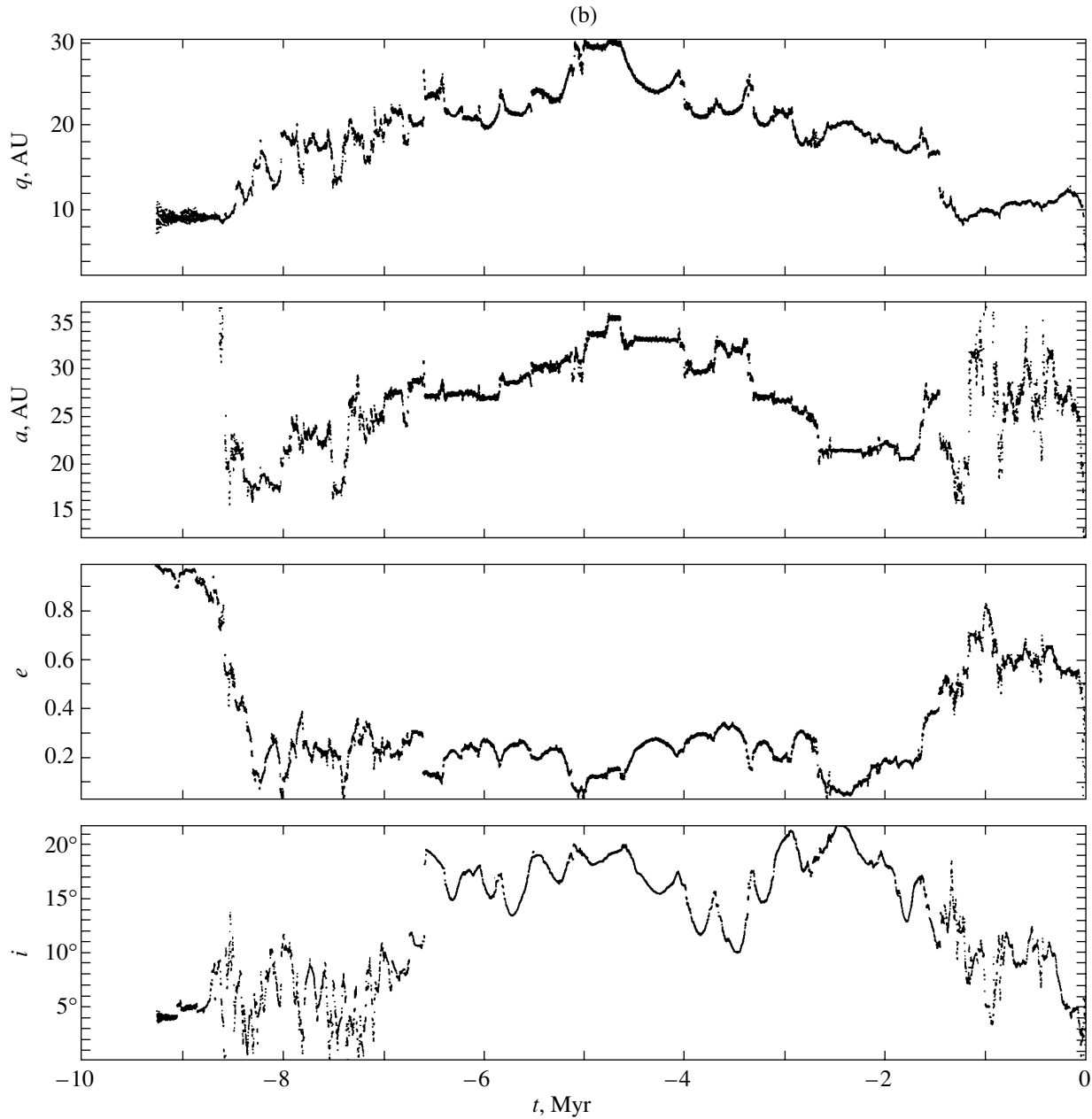

**Fig. 1.** Contd.

−0.4 and 0.6 Myr, and −0.8 and 0.2 Myr, respectively. For the RMVS3 integrator, this time was −0.45 and 0.3 Myr, and −0.25 and 0.5 Myr, respectively. Among the 48 absolute values of $T_h$ obtained with the use of the BULSTO integrator, six were smaller than 0.1 Myr, three exceeded 2 Myr, and eight were between 1 and 2 Myr.

When using the BULSTO integrator, in 10 of the 48 cases (20%) we obtained the minimum distance from the Sun, $q_{min} < 1$ AU (see table c, d). In the case of the integration to the future, this inequality was satisfied for 8 of the 24 runs (33%). Since by the time of the beginning of integration to the future an object has already crossed for some time Jupiter's orbit, our calcu-lations show that no less than 33% (up to 40%) of similar objects reached the Earth's orbit in the perihelion in the course of evolution. When using the symplex RMVS3 integrator, the number of the objects that reached the Earth's orbit was smaller by a factor of two (table e, f): 3 of the 24 objects for integration to the future and 1 of the 24 objects for integration to the past. This difference may be partly due to the fact that in the case of using the RMVS3 integrator, the step of calculations of the orbital elements (500–1000 yr) was several times larger than that for the use of the BULSTO integrator, and was comparable to the shortest time intervals, during which the Earth's orbit was crossed.

(c)

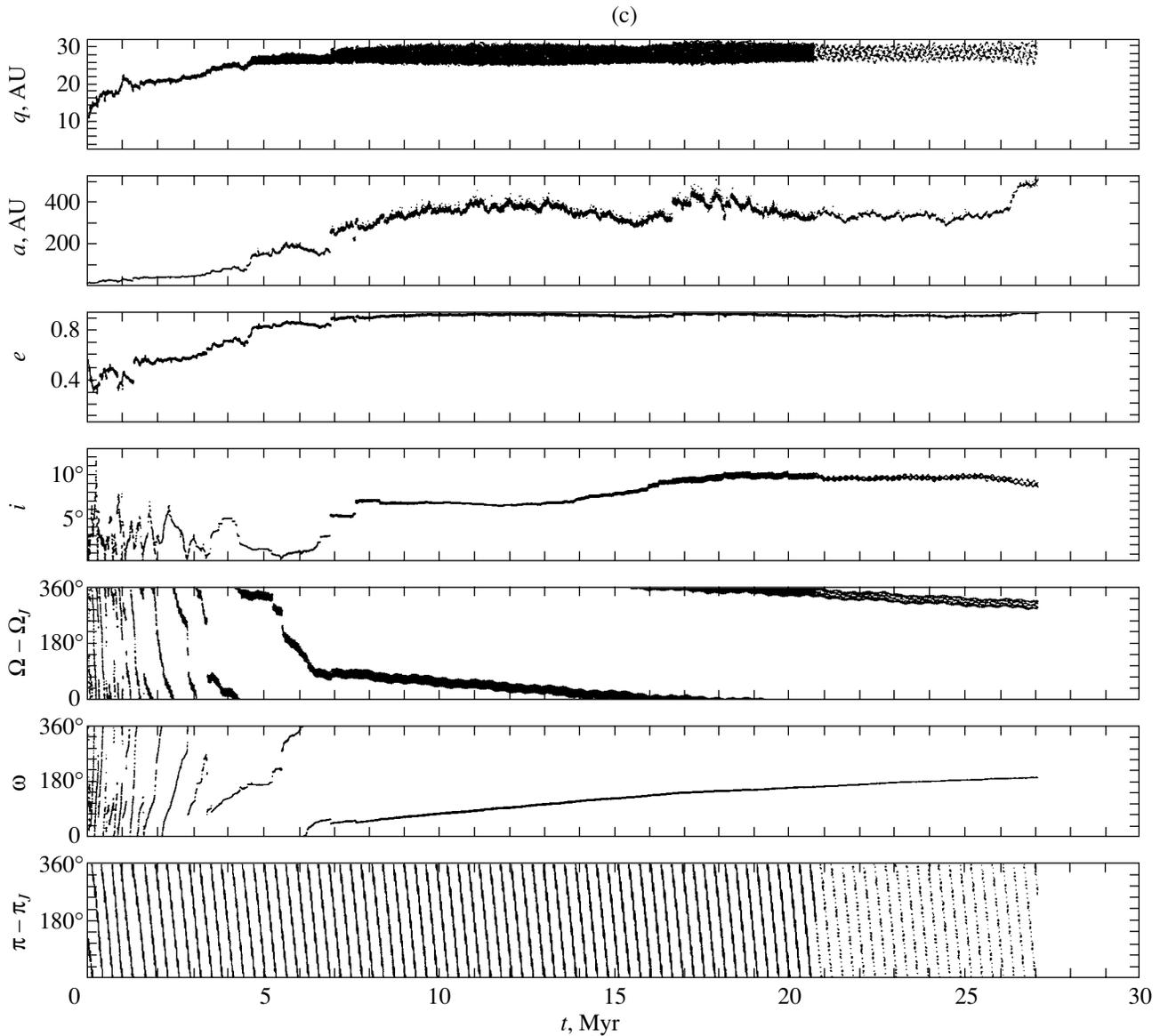

**Fig. 1.** Contd.

The mean value of the time interval, during which the object crossed the Earth's orbit during its lifetime, was about 5000 yr, but some intervals were several times smaller. Another (more important) reason for the above difference is that, in the case of using the RMVS3 integrator, bodies seldom get into resonances with planets, and the resonances with Jupiter sometimes sharply increase the orbital eccentricities of bodies and decrease their perihelion distances. Therefore, while investigating the evolution of the orbits of Jupiter- or Earth-crossing bodies, one must be careful in using the results obtained with the RMVS3 integrator.

For about 2/3 of the runs, at $q_{min} < 1$ AU, the minimum values of the semimajor axis $a$ were smaller than 3 AU, but at the same time they exceeded 2.75 and 2.62 AU for the use of the BULSTO and RMVS3 integrators, respectively. Among the 169 Apollo-asteroids known in March 1995, 10 asteroids had an $a$ between 2.5 and 2.75 AU, and there was only one asteroid for each of the intervals: $a = (2.75–3)$, $(3–3.5)$, $(3.5–4)$, and $a > 4$ AU. The semimajor axes of 14, 7, 3, 1, and 1 Amor-asteroids (out of 132) were in the same intervals, i.e., only for 8% of the Apollo asteroids and 20% of the Amor-asteroids $a > 2.5$ AU. Because the detection of asteroids with larger $a$ is more difficult, the portions of asteroids with such semimajor axes must be larger for all asteroids of the Apollo and Amor groups (not only for the already observed ones). In our calculations of the evolution of the orbits of objects close to the P/1996 R2 object's orbit, we did not obtain values

(d)

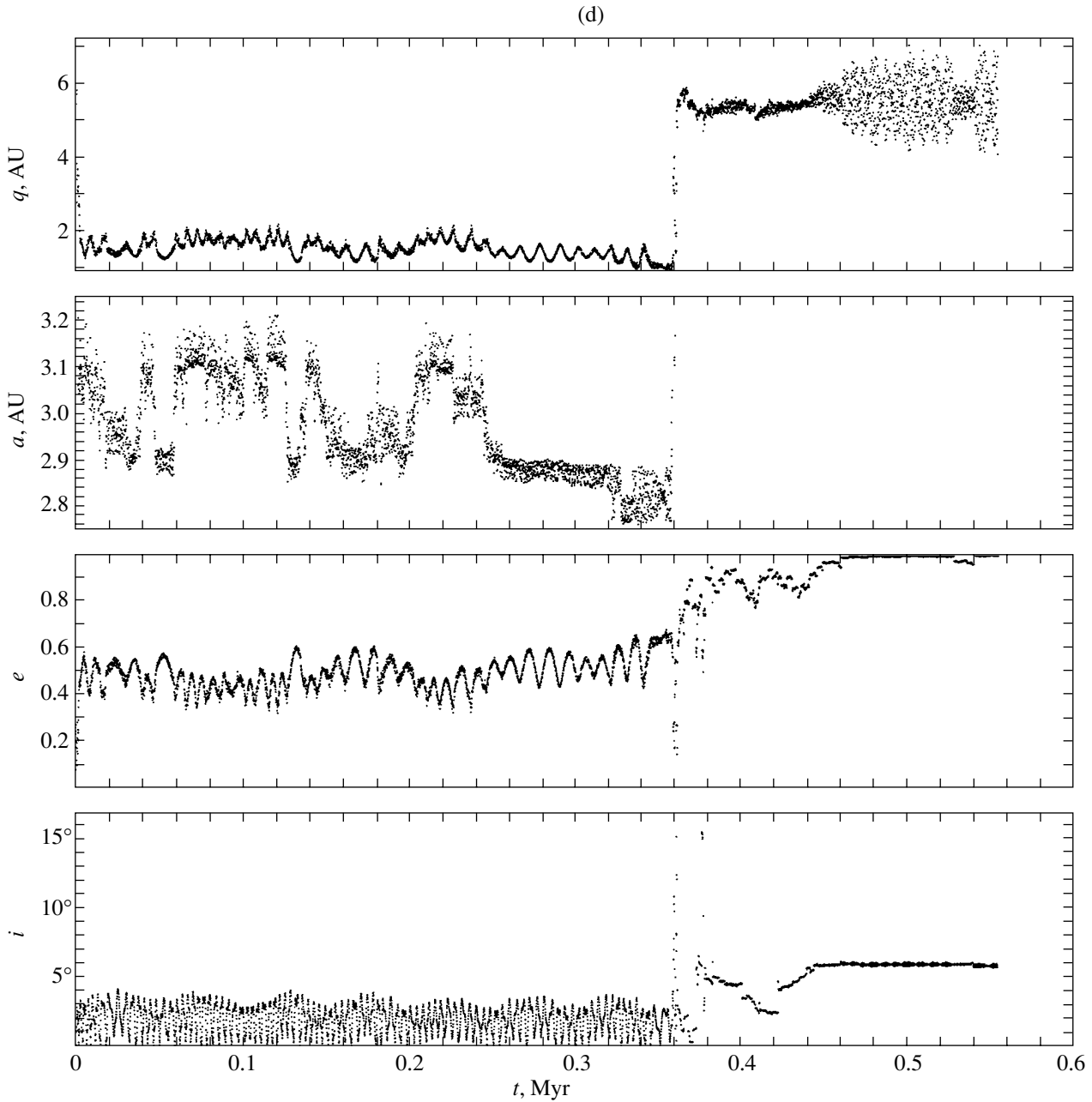

**Fig. 1.** Contd.

of $a < 2.5$ AU, which are typical for most of the Earth-approaching objects. However, while investigating the evolution of Jupiter-crossing orbits of bodies by the method of spheres of action, Ipatov (1995) obtained many bodies with $a < 2.5$ AU (and some bodies even with aphelia inside the Earth's orbit). By numerical integration of the equations of motion, we plan to investigate the evolution of orbits close to the orbits of other Jupiter-crossers in order to evaluate the probability of a considerable decrease in $a$ during evolution and to obtain numerical estimates of the portion of objects reaching the Earth's orbit in these cases.

The portion of the objects ejected into the orbits with eccentricities in the range $1 < e < 1.01$ (among the orbits with $e > 1$) was approximately 3/20 and 1/5 for the use of the BULSTO and RMVS3 integrators, respectively. In one of the runs, we obtained $e \approx 1.0001$ for both integrators. The bodies leaving the planetary system in near-parabolic orbits enter the Oort cloud.

In some runs, during its lifetime an object took part in less than 10 close encounters (within the radius of the Hill sphere) with the giant planets, and in other runs in several hundreds of close encounters. In total, in the 48 runs carried out with the RMVS3 integrator, there

(e)

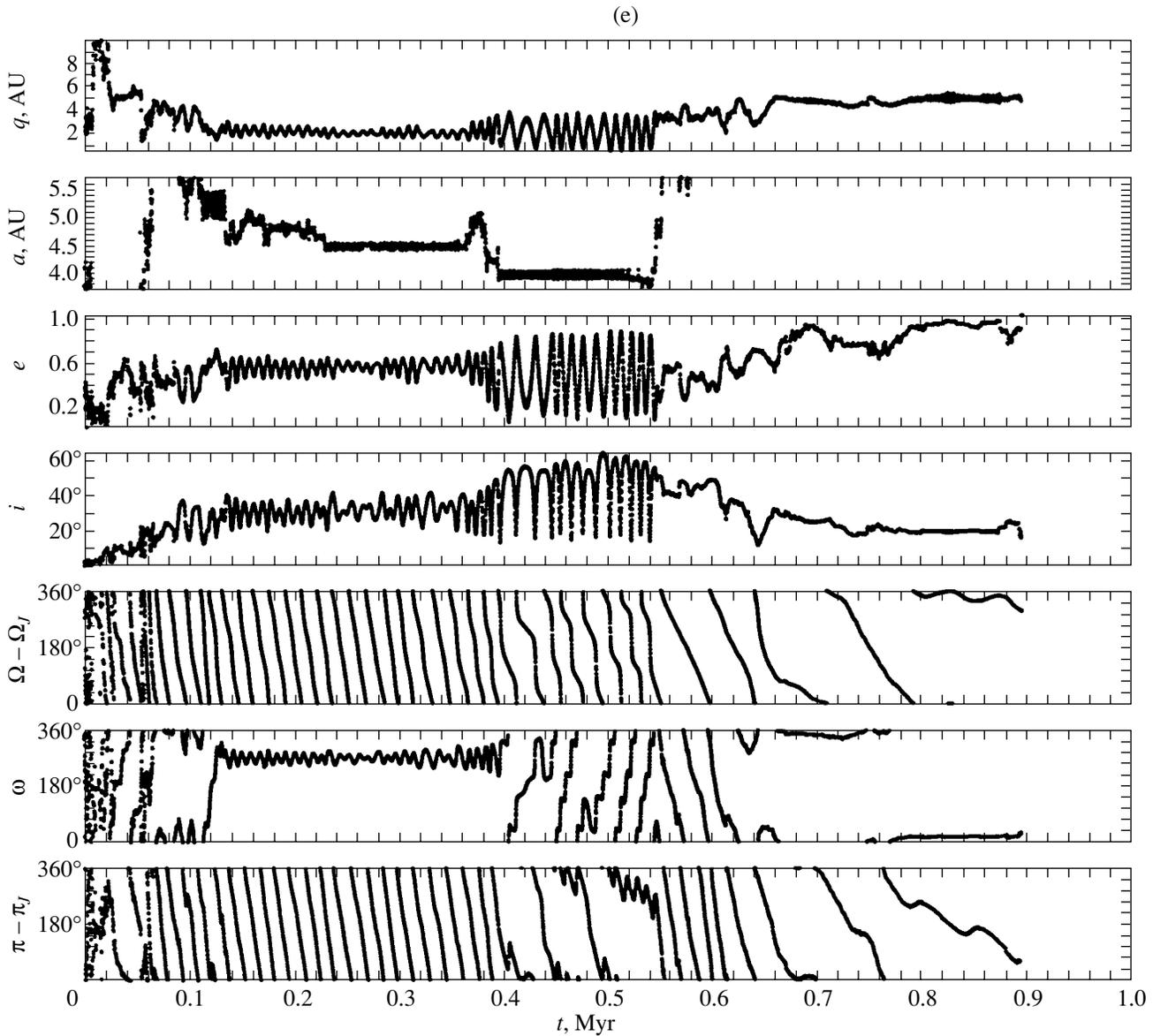

**Fig. 1.** Contd.

were 2469, 1598, 402, and 435 close encounters with Jupiter, Saturn, Uranus, and Neptune, respectively. Note for comparison that for these planets the squares of the ratios of the radius of the Hill sphere to the radius of a planet are 5300, 11360, 74450, and 217500, respectively. Thus, the probability of a collision of such objects with the giant planets is small and decreases with the distance of a planet from the Sun.

The plots of the time variations in the elements of the orbits close to the orbit of the P/1996 R2 object are presented in Fig. 1. Small variations in the initial orbits can cause considerable variations in the character of evolution. Usually $i < 20°$, including the moment of an ejection into a hyperbolic orbit. In Figs. 1–2, the orbital inclinations are presented relative to the initial (at $t = 0$) orbital plane of the Earth.

In the variant presented in Fig. 1a, the minimum distance of an object from the Sun 80 000 yr ago was 0.26 AU, i.e., a body entered inside the orbit of Mercury. About 5 Myr ago, another test object for some time moved in a weakly eccentric orbit in the trans-Neptunian belt (Fig. 1b). An example of the long lifetime (27 Myr) of an object is presented in Fig. 1c. In this case, the object went to a highly eccentric orbit with $a > 200$ AU, whose perihelion lay near the orbit of Neptune during about 20 Myr. Some actual celestial objects can also have such orbits.

(f)

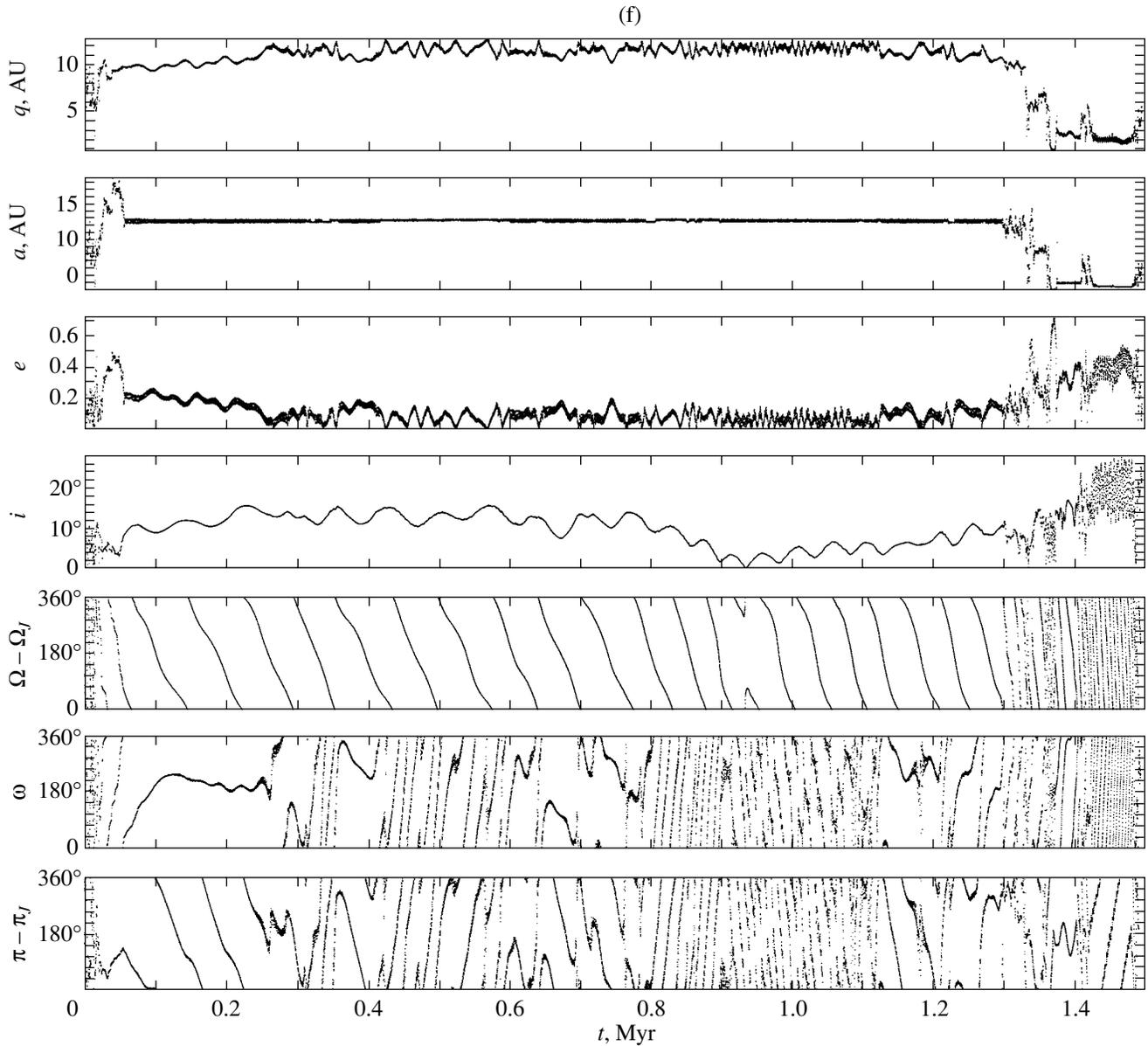

**Fig. 1.** Contd.

In contrast to Figs. 1a–1c, in Figs. 1d–1j the results of runs obtained with the use of the BULSTO integrator, but not the RMVS3 integrator, are presented. The use of a more accurate method of integration (BULSTO) allows one to better investigate the ejections of objects into resonant orbits. In the case presented in Fig. 1d, an object was in the 7 : 3 resonance with the motion of Jupiter ($a = 2.96$ AU) for some time, then got into the 5 : 2 resonance ($a = 2.82$ AU) where it reached the Earth's orbit in the perihelion, and after leaving this resonance sharply increased $a$ and the perihelion distance $q = a(1 − e)$. Hereafter we consider the ratios 5 : 2 and 7 : 3, most widely accepted throughout the world. In the USSR and Russia, the inverse ratios (2 : 5 and 3 : 7) were usually considered. Another test object (Fig. 1e) was in resonant orbits (resonances 1 : 1, 5 : 4, and 3 : 2 with the motion of Jupiter) for a longer time, being simultaneously in the 5 : 4 resonance and in the Kozai resonance ($\omega \approx 270°$). For the 3 : 2 resonance, the doubled amplitude of variations in $e$ reached 0.8, the doubled amplitude of variations in $i$ reached 40°, and, therewith, the object's orbit periodically crossed the Earth's orbit. The object was in the 2 : 3 resonance with the motion of Saturn during more than one million years in the case presented in Fig. 1f, and during 0.5 Myr in the case shown in Fig. 1g. For this resonance, variations in eccentricity did not exceed 0.2, as a rule. In the case of Fig. 1g, the semimajor axis of the orbit of the object was close for some time to the semimajor axis of Uranus's orbit. The case when the object was in the 2 : 1

(g)

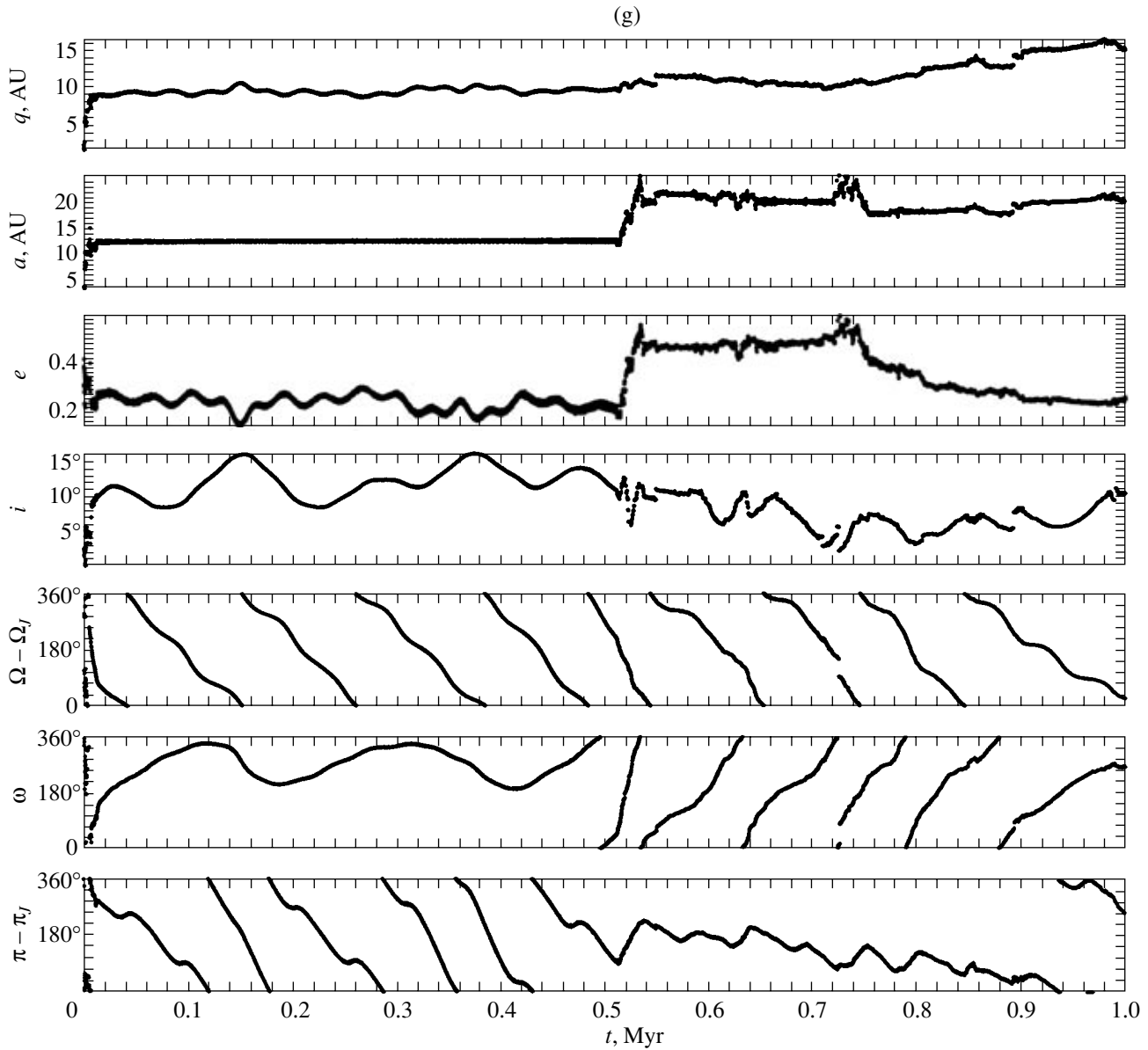

**Fig. 1.** Contd.

resonance with Jupiter ($a = 3.28$ AU) during 300 000 yr is presented in Fig. 1h. Therewith, the amplitude of $a$ was ~0.1 AU, and variations in $e$ and $i$ were relatively small.

In many cases, before the ejection of the object into a hyperbolic orbit (Figs. 1c, 1d, 1h for integration to the future and Figs. 1a, 1b, 1j for integration to the past), the eccentricity exceeded 0.9 for a relatively long time interval (and was sometimes close to unity). In these cases, the orbital perihelion usually was close to the orbit of a giant planet (Jupiter in Figs. 1a, 1d, 1j, Saturn in Figs. 1b, 1h, and Neptune in Fig. 1c). The objects were not ejected into hyperbolic orbits during the time intervals presented in Figs. 1f, 1g, 1i. The number of the variants considered, in which the orbital perihelion of an object was for some time near Jupiter's orbit before its ejection into a hyperbolic orbit, was greater by a factor of 4 than the number of variants, in which the perihelion was close to the orbit of Saturn. As a rule, Jupiter ejected an object relatively quickly, and the portions of objects ejected by Jupiter and Saturn in more than a million years are close. Only in one run, the orbital perihelion before the ejection of an object was close to the Neptune's orbit, and there were no cases when the perihelion was near the orbit of Uranus. In some runs, the orbital perihelion of an object before its ejection was far from the semimajor axes of any planets. For example, $q$ varied for a long time near 8 AU, or between 6 and 9 AU.

(h)

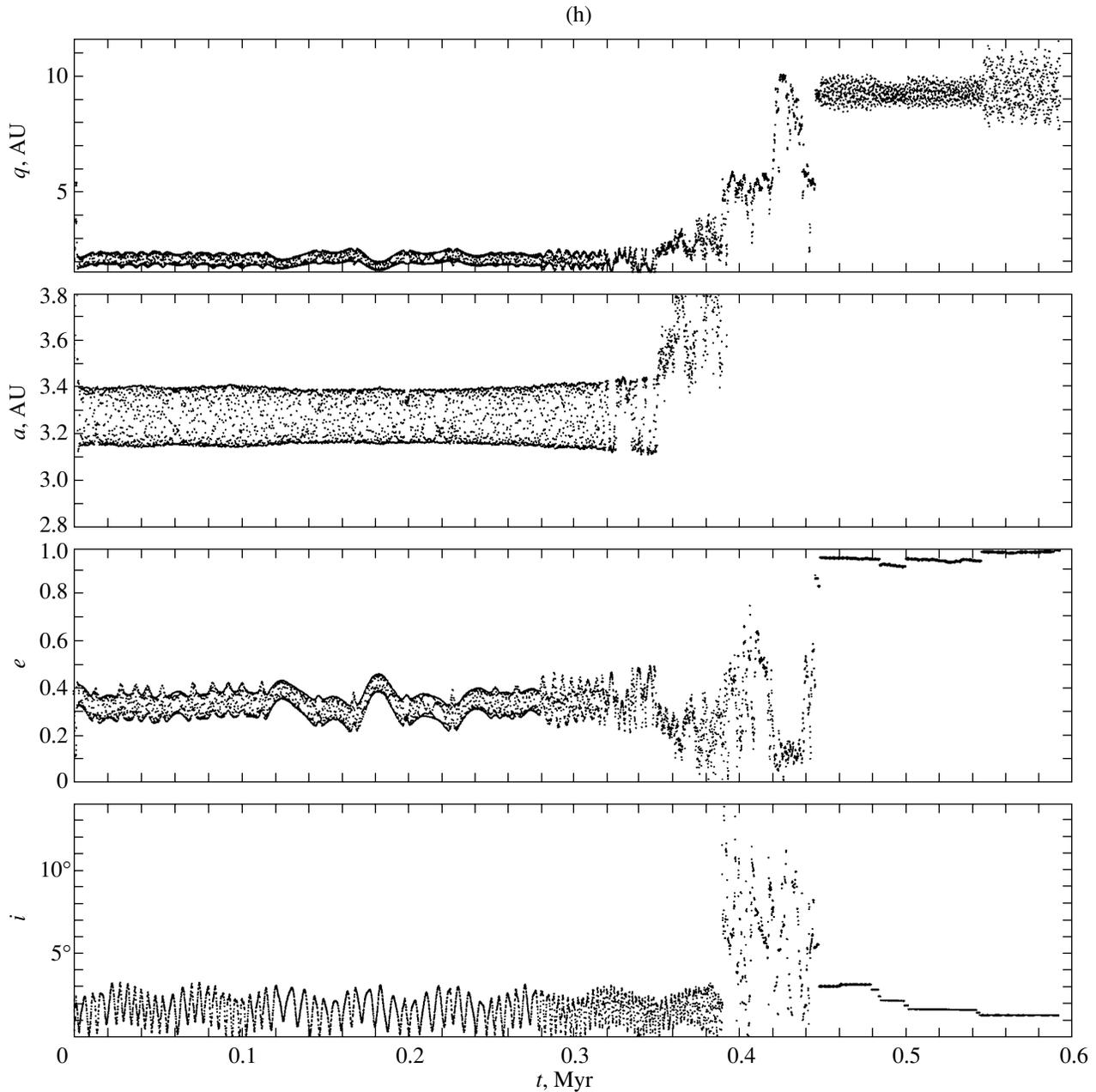

**Fig. 1.** Contd.

For the orbital evolution presented in Fig. 1i, the perihelion distance was repeatedly smaller than the semimajor axis of Mercury's orbit and reached 0.002 AU. At the time in the range of $-0.8$ Myr $< t < -0.55$ Myr, the object was in the 4 : 3 resonance with the motion of Jupiter ($a = 4.29$ AU), and during a major part of this time interval, was simultaneously in the Kozai resonance ($\omega \approx 270°$). Variations in $e$ and $i$ were large, and $i$ even reached 160°. At $t \sim 0.3$ Myr, during 40 000 yr the object was in the 1 : 1 resonance with Jupiter. The temporary captures of objects into this resonance, obtained by us, show that some of Jupiter's Trojans could be captured from outside.

## EVOLUTION OF ORBITS CLOSE TO THE P/1996 N2 ORBIT

Orbital elements of all orbits close to the orbit of the P/1996 N2 object varied quasi-periodically over the time interval considered (up to 200 Myr). The ranges of the variations in the semimajor axis, eccentricity, and inclination of this object were, respectively, $\Delta a = a_{max} - a_{min} \approx 0.041$ AU, $\Delta e \approx 0.11$, and $\Delta i \approx 3.5°$. For close

(i)

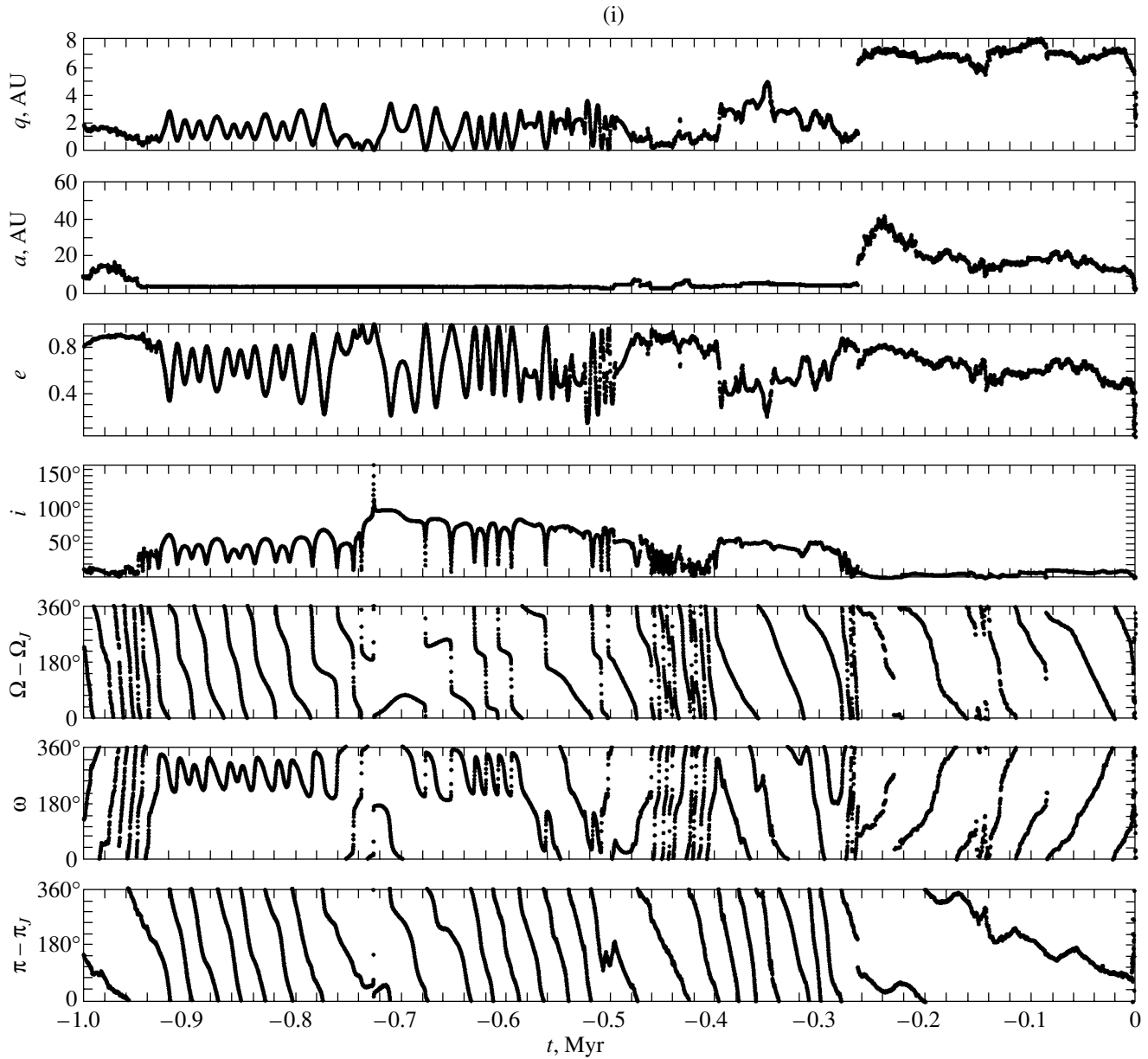

**Fig. 1.** Contd.

orbits, the differences in the values of $\Delta a$ and $\Delta e$ did not exceed 11% for variations in the initial $a$ values within $\pm 0.005$ AU, and the differences in the values of $\Delta a$ were 7% for variations in the initial eccentricities ($\pm 0.015$). For variations in the initial values of the inclination $i$ ($\pm 0.05°$), the ascending-node longitude $\Omega$ ($\pm 2°$), and the perihelion argument $\omega$ ($\pm 2°$), variations in $\Delta a$, $\Delta e$, and $\Delta i$ did not exceed 2%. Variations in the orbital elements of the P/1996 N2 object over a time interval of 1 Myr for integration to the past are presented in Fig. 2. The values of $\Delta a$, $\Delta e$, and $\Delta i$ obtained with the use of the RMVS3 integrator over time intervals equal to 20 and 200 Myr differ by no more than several percent from the corresponding values obtained with the use of the BULSTO integrator on the 1 Myr interval. In all the runs considered, $\omega$ increased by 360° over the time span $T_\omega \approx 6000$ yr, the values of $\Delta\Omega = \Omega - \Omega_J$ decreased by 360° over the time span $T_\Omega \approx 20000$ yr, and the values of $\Delta\pi = \pi - \pi_J$ increased by 360° over the time span $T_\pi \sim 10000$ yr, where $\pi = \Omega + \omega$, $\Omega_J$ and $\pi_J$ are the values of $\Omega$ and $\pi$ for Jupiter. The P/1996 N2 object has the smallest inclination among the actual asteroids with close $a$ values, and therefore the probability of its collision with other asteroids is larger.

For the P/1996 N2 object, the period of main variations in $a$ is 110 yr, and the amplitude of variations in $e$ with this period is close to 0.02–0.03. For such variations,

(j)

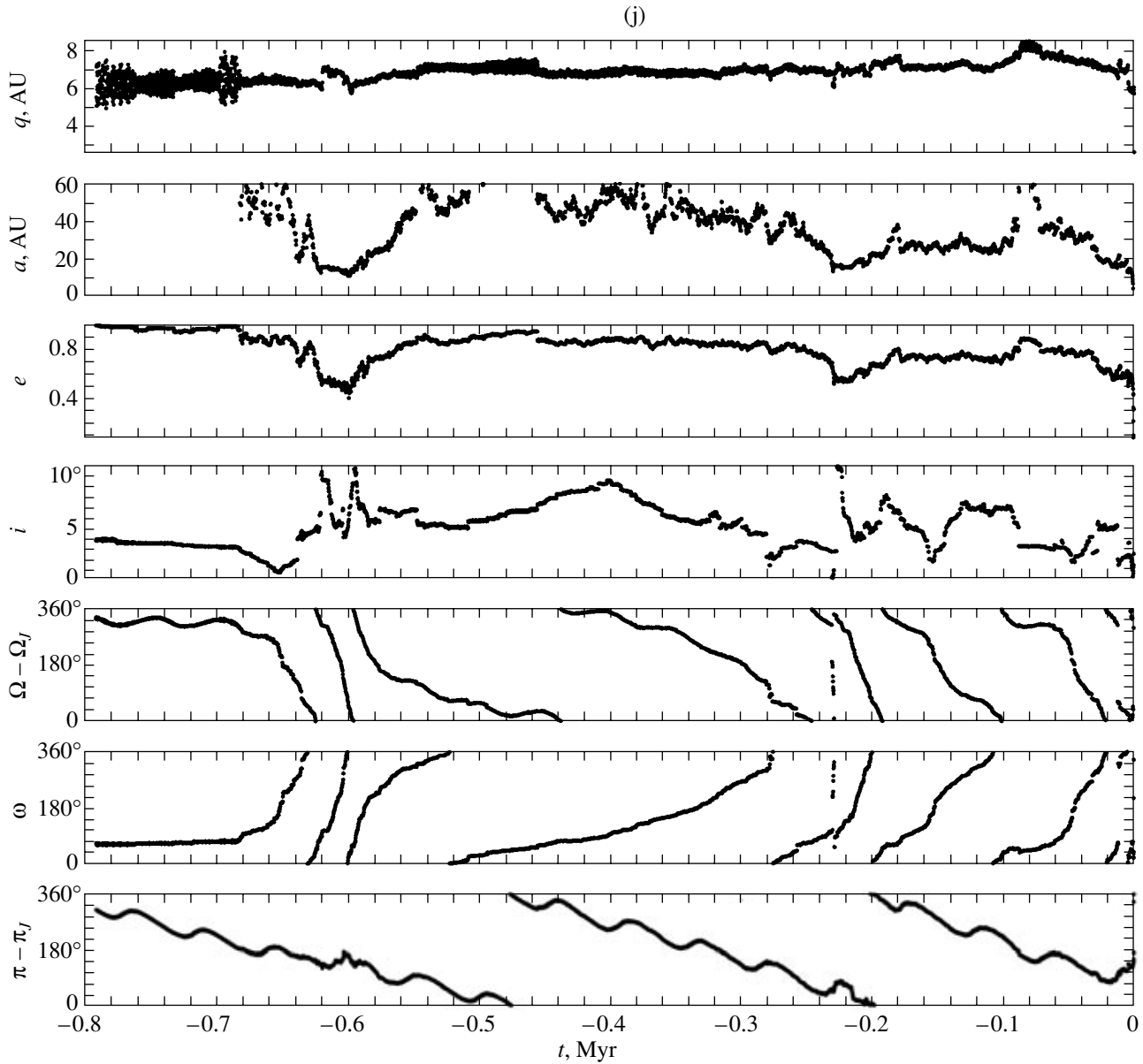

**Fig. 1.** Contd.

the minimum of $e$ corresponds to the maximum of $a$, and the maximum of $e$ corresponds to the minimum of $a$. The period of main variations in $e$ is $T_e \approx$ 8.4 thousand years, and $i$ varies with a period of $T_i \approx$ 12 thousand years. The values of $e$ and $i$ also vary with the periods $T_{eJ} \approx 54$ thousand years and $T_{iJ} \approx 49$ thousand years, respectively, where $T_{eJ}$ and $T_{iJ}$ are the periods of variations in $e$ and $i$ for Jupiter (the periods for Saturn are the same). The doubled amplitude of variations in $e$ with a period $T_e$, $2A = \max - \min$, varies from 0.06 to 0.1 during $T_{eJ}$. It has a maximum when the eccentricity $e_J$ of Jupiter reaches a maximum; the minimum $2A$ value corresponds to the minimum eccentricity $e_J$. All these variations in the amplitude are caused by the influence of Saturn. The variations in $e$ with the period $T_e$ are identical over the course of evolution, if we consider the three-body problem (the Sun–Jupiter–an asteroid).

As for most of the other nonresonant asteroids, $\Delta\pi$ for the P/1996 N2 object increases by 360° over $T_e$; $\Delta\pi \approx 0$ when $e$ reaches a maximum, and $\Delta\pi \approx 180°$ when $e$ reaches a minimum. The values of $i$ are smaller when the inclination $i_J$ of Jupiter has a minimum value. During $T_{iJ} \approx 4T_i$, one or two minimum values of $i$ are close to zero. At these instants of time, the values of $\Delta\Omega$ and $\omega$ change abruptly by 180°. During the remaining time, $\Delta\Omega$ oscillates about zero with the period $T_i$ and an amplitude that does not exceed 90°. When varying the initial values of $a$ and $e$ by ±0.005 AU and ±0.015, respectively, the variations in $T_\pi$ were ∓4 and ∓3%, and the variations in $T_i$ were ∓2 and ∓4%, respectively. For other variation in the initial data, the variations in $T_\pi$ and $T_i$ were much smaller.

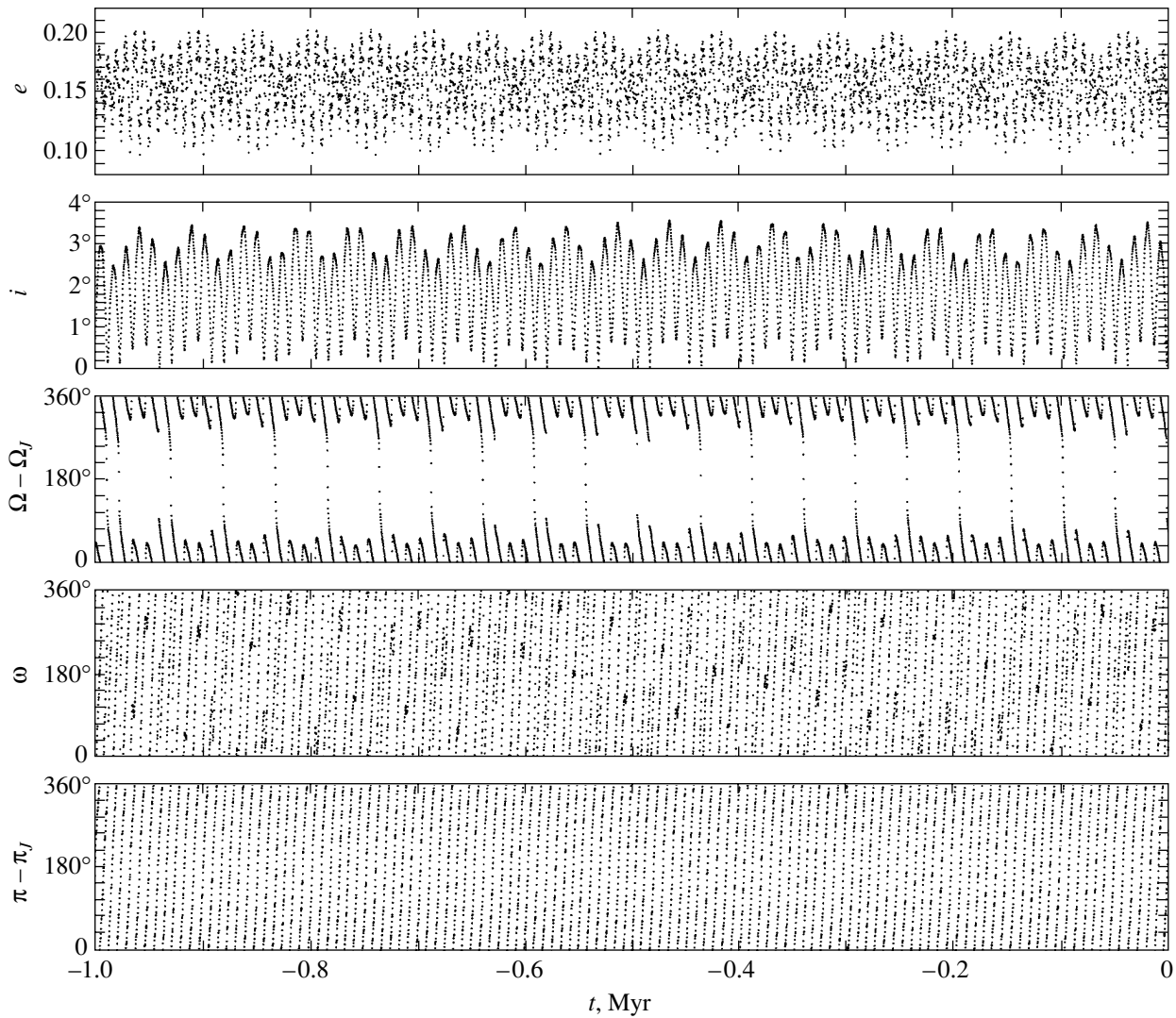

**Fig. 2.** Time variations (Myr) in the eccentricity $e$ and inclination $i$ (in degrees); the difference $\Delta\Omega = \Omega - \Omega_J$ in the ascending-node longitudes of the body and Jupiter and in the perihelion argument $\omega$; and the difference $\Delta\pi = \pi - \pi_J$ in the longitudes of the perihelion of the body and Jupiter (all angles are shown in degrees). The initial orbit is close to the orbit of the P/1996 N2 object. The results are obtained by numerical integration of the system (the Sun, planets, an object) with the use of the method by Bulirsh and Stoer.

## CONCLUSION

Small variations in the initial elements of the orbits of objects close to the orbit of the P/1996 R2 object can cause large variations in the character and limits of variations in the orbital elements. The characteristic lifetime (before the ejection of the object into a hyperbolic orbit) of such objects had values between 30 thousand and 27 million years. For our runs (based on the use of the symplex RMVS3 integrator and the usual BULSTO integrator), the median value of this characteristic lifetime was several hundreds of thousands years. In one of the considered runs for integration to the past, an object moved during some time in the trans-Neptunian belt in a slightly eccentric orbit. The main difference between the results obtained with the use of the RMVS3 and BULSTO integrators is that the objects considered move in resonances with planets more often when the BULSTO integrator is used. When using the latter integrator, during evolution, objects reached the Earth's orbit in about 1/4 of the considered runs (including 1/3 of the runs for integration into the future). In the case of the RMVS3 integrator, this number was about half as large.

Orbital elements of the P/1996 N2 object varied quasi-periodically over the considered time interval equal to 200 Myr. Plots of the variations in the orbital elements varied slightly for small variations in the initial orbits and also in using different integrators.

## ACKNOWLEDGMENTS

For S.I. Ipatov, this work was supported by the Russian Foundation for Basic Research, project code 96-02-17892, and by the Federal Scientific and Technical Program "Astronomy," Section 1.9.4.1. The runs made during the visit of S.I. Ipatov to Berlin in 1996 were supported by the DAAD grant (pkz A/96/20688).